\documentclass[fleqn,usenatbib]{mnras}
\usepackage[T1]{fontenc}
\usepackage{ragged2e,color}
\DeclareRobustCommand{\VAN}[3]{#2}
\RequirePackage{fix-cm}
\let\VANthebibliography\thebibliography
\def\thebibliography{\DeclareRobustCommand{\VAN}[3]{##3}\VANthebibliography}
\newcommand{\cii}{[C\,{\sc ii}]}

\newcommand{\oiii}{[O\,{\sc iii}]}

\newcommand{\oiiil}{[O\,{\sc iii}] 88\,$\mu{\rm m}$}
\newcommand{\ciil}{[C\,{\sc ii}] 158\,$\mu{\rm m}$}

\definecolor{referee}{RGB}{0,0,0}

\usepackage{graphicx}	
\usepackage{amsmath}	
\usepackage{amssymb}	
\usepackage{newtxtext,newtxmath}

\title[A warm ULIRG at $z = 8.3$]{A warm ultra-luminous infrared galaxy just 600 million years \\after the Big Bang}
\author[Bakx et al.]{T. J. L. C. Bakx$^{1}$\thanks{E-mail: tom.bakx@chalmers.se}, 
Laura Sommovigo$^{2}$, 
Yoichi Tamura$^{3}$, 
Renske Smit$^{4}$, 
Andrea Ferrara$^{5}$, 
Hiddo Algera$^{6}$, \newauthor{}
Susanne Aalto$^{1}$, 
Duncan Bossion$^{7}$, 
Stefano Carniani$^{5}$, 
Clarke Esmerian$^{1}$, 
Masato Hagimoto$^{3}$, \newauthor{}
Takuya Hashimoto$^{8,9}$, 
Bunyo Hatsukade$^{10,11,12}$, 
Edo Ibar$^{13,14}$, 
Hanae Inami$^{15}$, 
Akio K. Inoue$^{16,17}$, \newauthor{}
Kirsten Knudsen$^{1}$, 
Nicolas Laporte$^{18}$, 
Ken Mawatari$^{17,19}$,
Juan Molina$^{13,14}$, 
Gunnar Nyman$^{20}$, \newauthor{}
Takashi Okamoto$^{21}$, 
Andrea Pallottini$^{5,22}$, 
W. M. C. Sameera$^{1}$, 
Hideki Umehata$^{23,3}$, 
Wouter Vlemmings$^1$, \newauthor{} and
Naoki Yoshida$^{12}$. 
\\
$^{1}$Department of Space, Earth and Environment, Chalmers University of Technology, SE-412 96 Gothenburg, Sweden \\
$^{2}$Center for Computational Astrophysics, Flatiron Institute, 162 5th Avenue, New York, NY 10010, USA\\
$^{3}$Department of Physics, Graduate School of Science, Nagoya University, Aichi 464-8602, Japan \\ 
$^{4}$Astrophysics Research Institute, Liverpool John Moores University, 146 Brownlow Hill, Liverpool L3 5RF, UK\\
$^5$Scuola Normale Superiore, Piazza dei Cavalieri 7, I-56126 Pisa, Italy\\
$^6$Institute of Astronomy and Astrophysics, Academia Sinica, 11F of Astronomy-Mathematics Building, No.1, Sec. 4, Roosevelt Rd, Taipei 106319, Taiwan, R.O.C. \\
$^7$Institute of Physics of Rennes, UMR-CNRS 6251, University of Rennes, 35000 Rennes, France\\
$^{8}$Division of Physics, Faculty of Pure and Applied Sciences, University of Tsukuba, Tsukuba, Ibaraki 305-8571, Japan\\
$^{9}$Tomonaga Center for the History of the Universe (TCHoU), Faculty of Pure and Applied Sciences, University of Tsukuba, Tsukuba, Ibaraki 305-8571, Japan\\
$^{10}$National Astronomical Observatory of Japan, 2-21-1 Osawa, Mitaka, Tokyo 181-8588, Japan\\
$^{11}$Department of Astronomical Science, The Graduate University for Advanced Studies, SOKENDAI, 2-21-1 Osawa, Mitaka, Tokyo 181-8588, Japan\\
$^{12}$Institute of Astronomy, Graduate School of Science, The University of Tokyo, 2-21-1 Osawa, Mitaka, Tokyo 181-0015, Japan\\
$^{13}$Instituto de F\'{i}sica y Astronom\'{i}a, Universidad de Valpara\'{i}so, Avda. Gran Breta\~{n}a 1111, Valpara\'{i}so, Chile\\
$^{14}$Millenium Nucleus for Galaxies (MINGAL), Avda. Gran Breta\~{n}a 1111, Valpara\'{i}so, Chile\\
$^{15}$Hiroshima Astrophysical Science Center, Hiroshima University, 1-3-1 Kagamiyama, Higashi-Hiroshima, Hiroshima 739-8526, Japan \\
$^{16}$Department of Physics, School of Advanced Science and Engineering, Faculty of Science and Engineering, Waseda University, 3-4-1 Okubo, \\Shinjuku, Tokyo 169-8555, Japan\\
$^{17}$Waseda Research Institute for Science and Engineering, Faculty of Science and Engineering, Waseda University, 3-4-1 Okubo, Shinjuku, Tokyo 169-8555, Japan\\
$^{18}$Aix Marseille Université, CNRS, CNES, LAM (Laboratoire d’Astrophysique de Marseille), UMR 7326, 13388 Marseille, France\\
$^{19}$Department of Pure and Applied Physics, School of Advanced Science and Engineering, Faculty of Science and Engineering, Waseda University, 3-4-1 Okubo, \\Shinjuku, Tokyo 169-8555, Japan\\
$^{20}$Department of Chemistry and Molecular Biology, University of Gothenburg, SE 413 90, Gothenburg, Sweden\\
$^{21}$Department of Cosmosciences, Graduate School of Science, Hokkaido University, N10W8, Kitaku, Sapporo 060-0810, Japan\\
$^{22}${Dipartimento di Fisica ``Enrico Fermi'', Universit\'{a} di Pisa, Largo Bruno Pontecorvo 3, Pisa I-56127, Italy} \\
$^{23}$Institute for Advanced Research, Nagoya University, Furocho, Chikusa, Nagoya 464-8602, Japan
}

\date{Accepted 2025 October 04. Received 2025 September 30; in original form 2025 August 15}

\pubyear{2025}

\begin{document}
\label{firstpage}
\pagerange{\pageref{firstpage}--\pageref{lastpage}}
\maketitle

\begin{abstract}
We present an Atacama Large Millimeter/submillimeter Array (ALMA) Band 9 continuum detection ($3.3 \sigma$) of MACS0416\_Y1 that confirms the suspected warm dust (91$^{+62}_{-35}$~K) of this Lyman-Break Galaxy (LBG) at $z = 8.3$ with $\log_{10} M_{\ast}/$M$_{\odot} = 9.0 \pm 0.1$. 
A modified black-body fit to the ALMA Bands 3 through 9 data of MACS0416\_Y1 finds an intrinsic infrared luminosity of 1.0$^{+1.8}_{-0.6} \times{} 10^{12}\ \mathrm{L_{\odot}}$, placing this UV-selected LBG in the regime of Ultra Luminous Infrared Galaxies (ULIRGs). 
Its luminous but modest dust reservoir (1.4$^{+1.3}_{-0.5} \times{} 10^{6}\ \mathrm{M_{\odot}}$) is co-spatial to regions with a UV-continuum slope $\beta_{\rm UV} \approx -1.5$ as seen by \textit{James Webb Space Telescope} (\textit{JWST}) imaging. Although this implies some dust obscuration, the 
\textit{JWST} photometry implies less obscured star formation than seen in the complete characterization by ALMA, implying some spatial separation of dust and stars on scales below 200~pc, i.e., smaller than those probed by \textit{JWST} and ALMA. This source is an extreme example of dust-obscured star formation contributing strongly to the cosmic build-up of stellar mass, which can only be revealed through direct and comprehensive observations in the (sub)mm regime. 
\end{abstract}




\section{Introduction}

\begin{figure*}
    \centering
    \includegraphics[width=\linewidth]{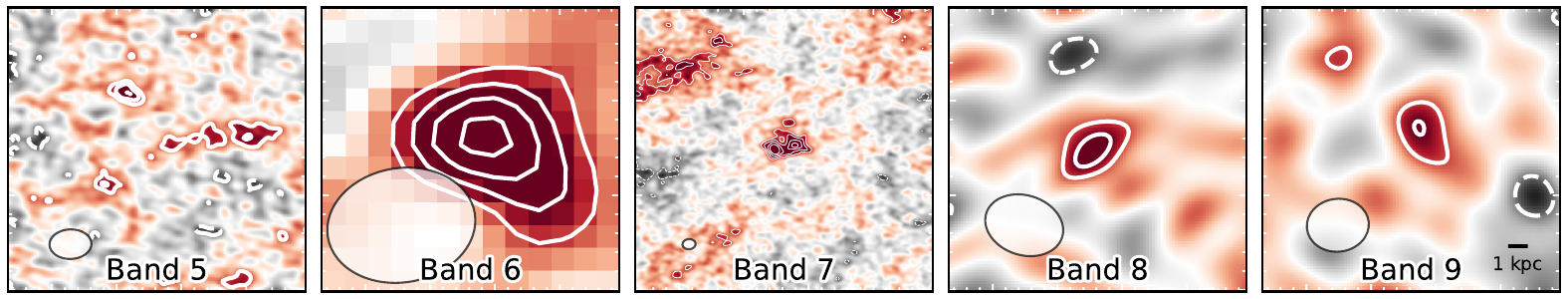}
    \caption{The dust continuum of MACS0416-Y1 in all (tentatively) detected Bands, namely 5, 6, 7, 8, and 9 -- covering rest-frame wavelengths 160 to 45$\mu$m. For Band 9, the 0.5~arcsec $uv$-tapered dust continuum is shown. The poststamps are 3 by 3 arcseconds, with the background image showing the continuum overlaid, together with \textit{white contours} indicating the $2, 3, 4, 5 \sigma$ emission. The beam is shown in the bottom-left corner, and the scalebar is shown in the bottom-right corner in the right-hand side poststamp. }
    \label{fig:continuum}
\end{figure*}
Observations with the Atacama Large Millimetre/submillimetre Array (ALMA) have revealed the presence of dust in galaxies in the Epoch of Reionization (e.g., \citealt{Watson2015,Tamura2019,Tamura2023,Bakx2020CII,Fudamoto2021OpticallyDark,Inami2022,Algera2024}). This was somewhat surprising, since ultraviolet (UV) studies that map out the unobscured star-formation rate density (SFRD) to $z \sim 10$ suggested a lack of dust at the high-redshift end based on the blue UV slopes of low stellar mass high-$z$ galaxies ($\beta_{\rm UV}$; e.g., \citealt{Finkelstein2015,Bouwens2015,Saxena2024,Ferrara2025EscapeFraction}). Although the \textit{James Webb Space Telescope} (\textit{JWST}) is finding reddened UV-slopes in the $z > 8$ Universe \citep[e.g.,][]{Langeroodi2024}, uncertainties still remain regarding the fraction of and extent of dust-obscured star formation in the early Universe because of the lack of comprehensive $z \gtrsim 8$ dust emission studies. These high-redshift studies are inherently challenging, as they demand substantial observational time even for single-band detections; yet constructing reliable spectral energy distributions (SEDs) requires multi-band coverage. Regardless of strategy, such surveys remain biased toward the brightest sources \citep{Inoue2016,Hashimoto2019} or gravitationally lensed objects \citep{Watson2015,Knudsen2017,Bakx2021,Akins2022,Fujimoto2024}, typically those pre-selected in the ultraviolet \citep[c.f.,][]{Bakx2024QSOcompanions}.

Initially, the strong far infrared emission at $z>7$ revealed by ALMA observations was attributed to the presence of unexpectedly large dust masses in the observed high-$z$ galaxies \citep{Lesniewska2019,Pozzi2021}. This exciting indication of rapid dust accumulation in the early Universe has guided studies into the physics of grain production and processing in the inter-stellar medium (ISM) in the early Universe \citep[e.g.,][]{Choban2022,Choban2024,Hirashita2023,Ferrara2023,Narayanan2025}. However, these constraints require accurate dust mass measurements (see e.g., \citealt{Inoue2020,Sommovigo2020hotdustorigins,Esmerian2024}), as they strongly depend on the dust temperature and instead most measurements currently resort to using an ad-hoc temperature instead ($T_{\rm d} \sim 35-50$~K; e.g., \citealt{Bouwens2020}).

Both recent observations (e.g., \citealt{Schaerer2015,Faisst2017,Laporte2019,Bakx2020CII,Viero2022,Witstok2023Dust}) and theoretical studies (e.g., \citealt{Sommovigo2020hotdustorigins,Sommovigo2021}) indicate that this ad-hoc approach to dust temperatures might not be appropriate to study dust-obscured star formation in the early Universe, as dust temperatures between galaxies can vary strongly, affecting the estimated dust masses and obscured star-formation rates \citep{Behrens2018}. Observationally, two galaxies at $z > 7.5$ have deep dust upper limits at $\sim 160$~$\mu$m that contrast against bright $\sim 90$~$\mu$m emission, indicating the presence of warm dust in these optically-selected galaxies ($T_{\rm d} \gg 60\ \mathrm{K}$; \citealt{Laporte2019,Bakx2020CII}). Such high temperatures can alleviate the large dust mass requirements set by the observed infrared emission \citep{Lesniewska2019,Sommovigo2020hotdustorigins}. This dust-temperature dependency is very strong indeed, affecting the dust mass and dust-obscured star-formation rate estimates by the sixth power, as the luminosity scales as $ T_{\rm d}^{4+\beta_{\rm d}}$, where $\beta_{\rm d} \sim 2.0$. 

While rest-frame UV to near-infrared dust studies at high redshift find evidence of dust through dust attenuation slopes \citep{Fisher2025,Markov2023,Markov2025}, extinction bumps \citep{Witstok2023Natur} and poly-aromatic hydrocarbons \citep{Spilker2023}, the properties of this dust in the early Universe might be very different to those observed in local galaxies \citep[c.f.][]{Sommovigo2025Attenuation}. 
Instead, (sub)mm studies with ALMA have revealed a tendency towards the measurement of higher $\beta_{\rm d}$ in galaxies \citep{Kato2018,Algera2024R25,Witstok2023Dust} and quasars \citep{Tripodi2023ApJ...946L..45T}. 
On top of the well-understood effects of the cosmic microwave background radiation, which both heats the dust and reduces the contrast against which we observe the dust \citep{daCunha2013}, the emissive properties of dust grains at $z > 7$ appear to contrast against the picture painted by recent \textit{JWST} observations. 
Alternative to -- or together with -- a high dust temperature, galaxies in the early Universe show a variety of dust attenuation curves \citep{Markov2023,Markov2025,Fisher2025} resulting from different compositions \citep{Demyk2017A&A...600A.123D,Demyk2017A&A...606A..50D} or size distributions \citep{Ferrara2017stochasticHeating,McKinney2025,Narayanan2025}, perhaps influenced by grain coagulation or icy mantles on the surface of grains, particularly in the denser regions of the ISM (see the discussion in \citealt{smith2017}). 

A comprehensive characterisation of the dust emission of distant galaxies is key to understanding their role within galaxy evolution. In an effort to characterize the formation of stars across cosmic time, the discrepancy between the obscured and unobscured cosmic SFRD in the $z > 5$ Universe has become ever more egregious (e.g., \citealt{Novak2017,Gruppioni2020,Talia2021}). Particularly, the galaxies that dominate the sub-mm emission at cosmic noon ($z = 1 - 5$) are not yet abundant enough in the cosmic dawn, adding additional uncertainty to the high-redshift cosmic SFRD. 
A population of low-metallicity, young galaxies with small but warm dust reservoirs could dominate the cosmic SFRD \citep{Sommovigo2022REBELS}, where even a small amount {\color{referee} of} dust is sufficient to obscure most of the galaxy. Such galaxies would be missed in studies that focus on $\sim 0.8 - 2$~mm detected galaxies, even from deep fields conducted with ALMA \citep{Zavala2021}, and could alleviate the discrepancy in cosmic star-formation rate estimates \citep{Viero2022}.


The galaxy at the centre of this study, MACS0416\_Y1 (Y1 hereafter, located at the IRCS coordinates 04:16:09.401, $-$24:05:35.47), could be an archetypal example of a low-metallicity galaxy with a small but warm dust component. Y1 is a $z = 8.31$ Lyman-break galaxy (LBG) with both robust \oiiil{} and \ciil{} line detections {\color{referee} as well as a} resolved dust continuum detection {\color{referee} at a resolution of 300 parsec}. The source was originally identified as a $z \approx 8$ bright ($H_{160} = 25.9$) LBG behind the \emph{Hubble} Frontier Field cluster MACS\,J0416.1$-$2403. Y1 is moderately magnified ($\mu_{\rm g} = 1.5$, \citealt{Kawamata2016,Rihtar2024arXiv240610332R}) with negligible differential lensing. The spectroscopic redshift was initially identified using the \oiiil{} line \citep{Tamura2019}.
The dust continuum at 90~$\mu$m, adjacent to the \oiii{}, was clearly detected at $> 5 \sigma$. However subsequent attempts to reveal the dust continuum at longer wavelengths did not detect any emission around the \ciil{} line \citep{Bakx2020CII}. These deep observations provide a strict upper limit at 160~$\mu$m, and indicate that the $90$~$\mu$m continuum detection still lies on the Rayleigh-Jeans part of the slope of the modified black-body emission, with a dust temperature in excess of $\sim 80$~K. 

Only high-frequency observations, probing close to or on the other side of the peak of the dust spectrum, can conclusively measure the dust temperature of a galaxy (e.g., \citealt{Bakx2021,Algera2024R25}). A study of the dust temperature using Band~8 at 70~$\mu$m finds a $3.5 \sigma$ detection at $254 \pm 81$~$\mu$Jy \citep{Harshan2024}, but did not probe close enough to the peak of the spectrum to put a lower bound on the dust temperature. 
This paper presents the high-frequency continuum measurements that are finally able to constrain the dust temperature in Y1. Combined with previous sub-mm results (Section~\ref{sec:obs}), we provide a detailed analysis of the dust spectrum (Section~\ref{sec:sed}) to investigate the properties of the furthest direct detection of dust emission in the Universe (Section~\ref{sec:implications}. We provide the conclusions in Section~\ref{sec:conclusions}. 
\footnote{Throughout this paper, we assume a flat $\Lambda$-CDM cosmology with the best-fit parameters derived from the \textit{Planck} results \citep{Planck2020}, which are $\Omega_\mathrm{m} = 0.315$, $\Omega_\mathrm{\Lambda} = 0.685$ and $h = 0.674$.}


\section{Observations}
\label{sec:obs}

\begin{table}
	\caption{Continuum measurements and results from the SED fits for MACS0416-Y1}
	\label{tab:data}
	\begin{tabular}{llll} 
\hline	
		& $\lambda$\,[mm] & $S_{\nu}^{\rm obs}$\,[$\mu$Jy] & Reference \\
\hline
Band 9 / 671 GHz & 0.447 & 465 $\pm$ 147 & This work \\
Band 8 / 465 GHz & 0.645 & 254 $\pm$ 81 & \citet{Harshan2024}  \\
Band 7 / 353 GHz & 0.849 & 137 $\pm$ 26 & \citet{Tamura2019,Tamura2023}   \\
\textit{Band 6} / 263 GHz & 1.14 & $<174$ (3$\sigma$) & \citet{Bakx2020CII} \\
Band 6 / 227 GHz & 1.31 & $41.7 \pm 6.5$ & This work\\
Band 5 / 197 GHz & 1.52 & 12 $\pm$ 6  & \citet{Bakx2020CII} \\
& & & Bakx et al. in prep.    \\
\textit{Band 4} / 143 GHz & 2.09 & $< 340$ ($3 \sigma$)  & This work    \\
\textit{Band 3} / \,\,\,93 GHz & 3.23 &  $< 21$ ($3 \sigma$)  & \citet{Jones2024}    \\
		\hline
		  \multicolumn{4}{c}{\textbf{SED fits}}  \\ 
		\hline \vspace{0.1cm}
$T_{\rm d}$ [K] 											& 91$^{+62}_{-35}$ 		& $72_{-9}^{+10}$ 		& $94_{-14}^{+17}$ 		\\ \vspace{0.1cm}
$\beta_{\rm d}$ 											& $1.52_{-0.75}^{+1.09}$ & 2 ({\it fixed}) 				& 1.5 ({\it fixed})  				\\ \vspace{0.1cm}
$M_{\rm d}$ [$10^{6}\ \mathrm{M_{\odot}}$]$^{\dagger}$ 		& 1.4$^{+1.3}_{-0.5}$ 	& $1.9^{+0.7}_{-0.5}$ 	& $1.4^{+0.5}_{-0.4}$ 		\\ \vspace{0.1cm} 
$L_{\rm IR}$ [$10^{12}\ \mathrm{L_{\odot}}$]$^{\dagger}$ 	& 1.0$^{+1.8}_{-0.6}$ 	& 0.7$^{+0.5}_{-0.3}$ 	& 1.1$^{+1.0}_{-0.5}$ 		\\ \vspace{0.1cm}
${\rm log_{10}\ IRX}$ 										& 1.4$^{+0.5}_{-0.3}$ 	& 1.3$^{+0.3}_{-0.3}$ 	& 1.4$^{+0.4}_{-0.3}$ 		\\ 
$M_{\rm d}$/SN [$\mathrm{M_{\odot}}$] 			& 0.07$^{+0.06}_{-0.03}$ 	& 0.05$^{+0.04}_{-0.03}$ 	& 0.06$^{+0.03}_{-0.02}$ 		\\
\hline
	\end{tabular}
	\raggedright \justify \vspace{-0.2cm}
\textbf{Notes:} $\,^{\dagger}$ Corrected for the magnification assuming $\mu = 1.5$ from \cite{Kawamata2016,Rihtar2024arXiv240610332R}, note that the listed fluxes are not corrected for lensing. Non-detections are listed in italics, while the $uv$-based extraction of the Band~5 is listed as a $2 \sigma$ tentative detection.
\end{table}
Using ALMA Band 9 \citep{Baryshev2015} observations at 671~GHz, we aim to measure the dust temperature of Y1 by characterizing its short-wavelength dust emission. Band~9, and specifically the 671~GHz window, provides the optimum frequency to measure Y1's likely warm dust temperature, with a more extensive justification given in Appendix~\ref{sec:whyband9}. In total, four observations accumulate a total of 4.9 hours including overheads between November 9$^{\rm th}$ and December $20^{\rm th}$ 2024 in the compact configurations C43-1, C43-2 and C43-3 with baseline lengths between 14 and 500~m (see Appendix Table~\ref{tab:observationdetails}). Two quasars, J0348–2749 and J0519-4546, were used for complex gain calibration, and another quasar, J0423-0120, was used for bandpass calibration. The data were calibrated using the pipeline script provided.

The continuum image is produced with \texttt{CASA} pipeline version 6.5.4.9 \citep{McMullin2007,TheCASATeam_2022} using natural weighting {\color{referee} to extract the continuum flux with the highest significance}. The dust continuum {\color{referee} emission is seen extended across 1.5 beams at 2$\sigma$. In an effort to best extract the flux, we aim to extract the flux directly from the peak pixel using tapered imaging. This flux extraction method tends to be more robust against noise peaks than aperture-based curve-of-growth algorithms, especially when the signal-to-noise ratio of the data is modest, see \citep{vanderVlugt2021,Algera2024R25}. As such,} we explore the total flux from the band 9 image through eleven different $uv$-tapering settings between 0 and 1 arcseconds at 0\farcs{}1 intervals. The total flux is estimated on each image through a curve-of-growth analysis, in an effort to extract the total available flux with a sufficiently low total noise estimate. The signal strength plateaus at a tapering of 0\farcs{}4, indicating that the source remains unresolved at this and larger taper values. 
Figure~\ref{fig:continuum} shows the resulting tapered Band~9 image with a 0.66 by 0.57 arcsecond beam with a position angle of 84 degrees, with an r.m.s. level of 140\,$\mu$Jy\,beam$^{-1}$ and a peak flux density of 465 $\pm$ 140\,$\mu$Jy ($\sim 3.3 \sigma$). An additional 10~per cent flux calibration uncertainty is added in quadrature, based on the recommendations from the ALMA technical handbook.\footnote{\url{https://almascience.eso.org/proposing/technical-handbook}}

Additionally, deep Band~6 data was taken to search for the [O\,{\sc i}]~146~$\mu$m emission (PID: 2024.1.00995.S; P.I. Renske Smit). The data are cleaned using the same procedure as described for the Band~9 data, i.e., using the pipeline script provided. Being a line detection experiment, the observations -- at a central frequency of 226.8~GHz -- were not designed to spatially resolve Y1. In imaging the Band~6 continuum emission of the source, we excluded channels potentially contaminated by the [O\,{\sc i}] line, defined as those within $\pm191\,\mathrm{km/s}$ -- the line FWHM of \cii{} in \citet{bakx2020} -- around a central redshift of $z=8.3116$. Adopting natural weighting, the image attains a sensitivity of $\sigma = 6.5\,\mu$Jy/beam, and has a beam size of 1\farcs{}6\,$\times$\,1\farcs{}2. At this resolution, Y1 remains unresolved and we measure a peak flux density of $41.7 \pm 6.5\,\mu\mathrm{Jy/beam}$ ($6.4\sigma$). 

The Band~4 data (2022.1.01356.S; P.I. Eiichi Egami) originates from an ALMA mapping observation. Using the online-reduced data from the ALMA science archive, we investigate the Band~4 emission at the position of Y1 using the primary-beam corrected continuum image. Y1 lies at the edge of the mapped region and remains undetected, with a primary-beam correction of 0.40. Using natural weighting, the $3\sigma$ flux limit is estimated to be $< 340$~$\mu$Jy, with a beam size of 1\farcs{}6\,$\times$\,1\farcs{}1.

The flux measurements of Y1 in Bands 3 through 9, including those from several shallower surveys, are provided in Table~\ref{tab:data}.\footnote{Following \cite{Bakx2024Receivers}, our observations are conducted using the Local Oscillator \citep{Bryerton2013} and the following receivers: The Band 3 \citep{Claude2008,Kerr2014}, Band 4  \citep{Asayama2014}, Band 5 \citep{Belitsky2018}, Band 6 \citep{Ediss2004,Kerr2004,Kerr2014}, Band 7 \citep{Mahieu2012}, Band 8 \citep{Sekimoto2008}, and Band 9 \citep{Baryshev2015}.} The images of the Bands in which Y1 is (tentatively) detected are shown in Figure~\ref{fig:continuum}. The higher-resolution data yields a strong 137~$\mu$Jy continuum detection at $90$~$\mu$m, with only an 18~$\mu$Jy upper limit at $3 \sigma$ after substantial tapering (i.e., smoothing the image in the $uv$-plane) at 160~$\mu$m. Using $uv$-based flux extraction with UVMULTIFIT \citep{MartiVidal2014}, a total flux $12 \pm 6$~$\mu$Jy is measured from the combined Band~5 data (Bakx et al. in prep.), in agreement with the flux limit of the tapered map. Based on the $uv$-based flux extraction of Band~5, we identify this emission as tentatively detected at $2 \sigma$. The subsequent analysis of fitting does not change appreciatively between a $2 \sigma$ detection and a $3 \sigma$ upper limit. All reported continuum fluxes are extracted from imaging that has been sufficiently tapered to leave the source unresolved.

\section{Fitting of the Spectral Energy Distribution} \label{sec:sed}
\begin{figure}
    \centering
    \includegraphics[width=\linewidth]{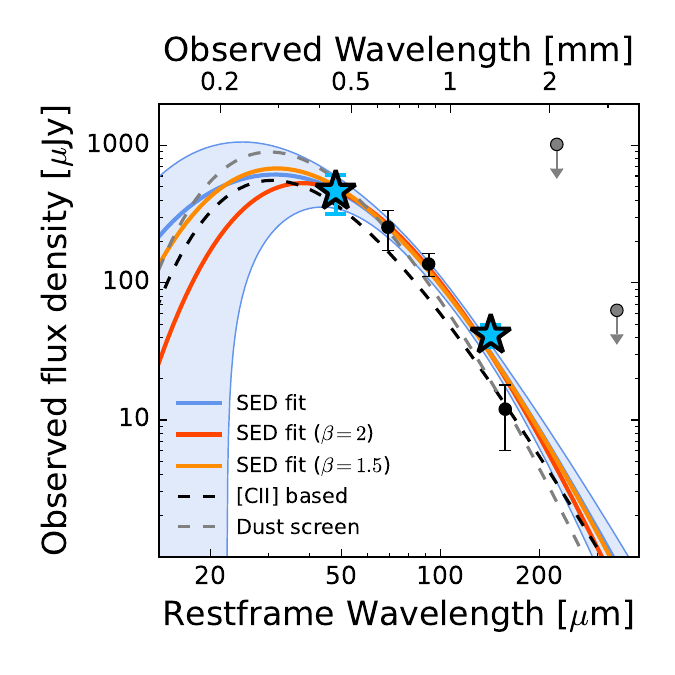}
    \caption{Three modified black-bodies are fitted to the data of Y1 (\textit{blue stars} for novel data, and \textit{black circles} for archival data, with upper limits in \textit{grey circles}), with different considerations of the $\beta_{\rm d}$ slope. The \textit{blue line and fill} indicates the best-fit results of a free $\beta_{\rm d}$ fit, while the \textit{red} and \textit{orange} lines indicate fits with fixed $\beta_{\rm d}$ of 2 and 1.5, respectively. The best-fit temperatures (listed in Table~\ref{tab:data}) range between 72 and 94~K. The \textit{dashed lines} show the new \cii{}-based dust temperature using the methodology of \citet{Sommovigo2021} in \textit{black} (see Appendix for details) and the dust scaling relation from \citet{Fudamoto2022DustTemperatures} \textit{grey}.}
    \label{fig:SEDfit}
\end{figure}

Figure\ \ref{fig:SEDfit} shows the modified black-body (eq. 8 in \citealt{Sommovigo2021}) fitted to the continuum points reported in Table~\ref{tab:data}. We use equations~12 and 18 from \cite{daCunha2013} to account for the heating of dust by and decreasing contrast against the CMB, respectively, with the $T_{{\rm CMB}, z = 8.3} = 25.4$~K. We approximate the dust mass absorption coefficient ($\kappa_{\rm \nu}$) as $\kappa_{\rm \star} \left( \nu / \nu_{\star}\right)^{\beta_{\rm d}}$, with ($\kappa_{\rm \star}, \nu_{\star})$ as (10.41\,cm$^2$/g, 1900\,GHz) from \cite{Draine03}.
We use the \texttt{emcee} MCMC-fitting routine \citep{ForemanMackey2013}, and allow $\log_{10} M_{\rm d}$, $T_{\rm d}$ and $\beta_{\rm d}$ to vary freely using flat priors, resulting in a dust mass of 1.4$^{+1.3}_{-0.5}$\,$\times{} 10^{6}$\,M$_{\odot}$, a dust temperature of 91$^{+62}_{-35}$\,K and a $\beta_{\rm d}$ of $1.52_{-0.75}^{+1.09}$. Most importantly, the demagnified infrared luminosity of Y1 is 1.0$^{+1.8}_{-0.6} \times{} 10^{12}\ \mathrm{L_{\odot}}$, indicative of an Ultra-Luminous Infrared Galaxy (ULIRG) at redshift 8.31. We explore the alternative to using a $12 \pm 6$~$\mu$Jy data point for the Band~5 flux density based on UVMULTIFIT \citep{MartiVidal2014} analysis, finding that the results do not change when taking a $3\sigma$ upper limit based on the $\chi^2$ error propagation as detailed in \cite{Sawicki2012}. {\color{referee} Using this $\chi^2$ error propagation, the exclusion of the two upper limits in Bands~3 and 4 do not significantly change the fitting results.}

We note that the spectrum appears well-represented by a single modified black-body, whereas before the availability of the Band 9 data, convergence of the fit was not possible with flat priors \citep{Bakx2020CII,Algera2024}. The only stand-out data point is from the Band~6 data, which has a much larger beam size compared to all other frequencies. Although little additional flux is seen in our curve-of-growth analysis on the other bands, this larger beam could include more extended dust missed by other observations, biasing the dust temperature and far-infrared luminosity to slightly lower values. Smoothing all data to the same resolution as Band~6 results in significantly larger uncertainties on the flux measurement, and cannot constrain the dust temperature measurement through a modified black-body fit. Excluding the Band~6 data provides a similar dust temperature ($88_{-31}^{+63}$~K), a steeper $\beta_{\rm d}$ ($ = 1.88_{-0.84}^{+1.04}$) and increases the infrared luminosity by roughly $\sim 10$ per cent, which falls within the current best-fit uncertainty.  Alternatively, if the Band~5 data is biased low, the resulting dust temperature fit excluding the Band~5 data results in much higher dust temperatures ($116_{-52}^{+46}$~K) and substantially lower $\beta_{\rm d}$ $(= 0.8_{-0.5}^{+0.9}$), as well as a 10~per cent higher infrared luminosity. 

If we take a fiducial $\beta_{\rm d} = 2.0$ or 1.5, we find a more accurate dust temperatures of $72_{-9}^{+10}$ and $94_{-14}^{+17}$\,K, respectively. However there is no appreciable improvement of the error of dust mass ($1.9^{+0.7}_{-0.5}$ and $1.4^{+0.5}_{-0.4} \times 10^6$\,M$_{\odot}$, resp.) nor luminosity, while finding ULIRG or ULIRG-like luminosities. We note that the assumption on a single dust temperature is a likely oversimplification \citep{Dunne2001,Behrens2018,Liang2019,DiMascia2021}. Simulation-based analytical models with an assumed log-normal distribution in dust temperatures predict that relatively wide distributions are necessary to strongly affect the total dust mass and infrared luminosity estimates \citep{Sommovigo2025}. Similar to the $z = 7.1$ galaxy A1689-zD1, the accuracy of these latter parameters thus depends solely on observational uncertainties, indicating that we fully trace the dust emission in this source.

\section{Implications}
\label{sec:implications}

\subsection{A ULIRG at redshift 8}

\begin{figure}
    \centering
    \includegraphics[width=\linewidth]{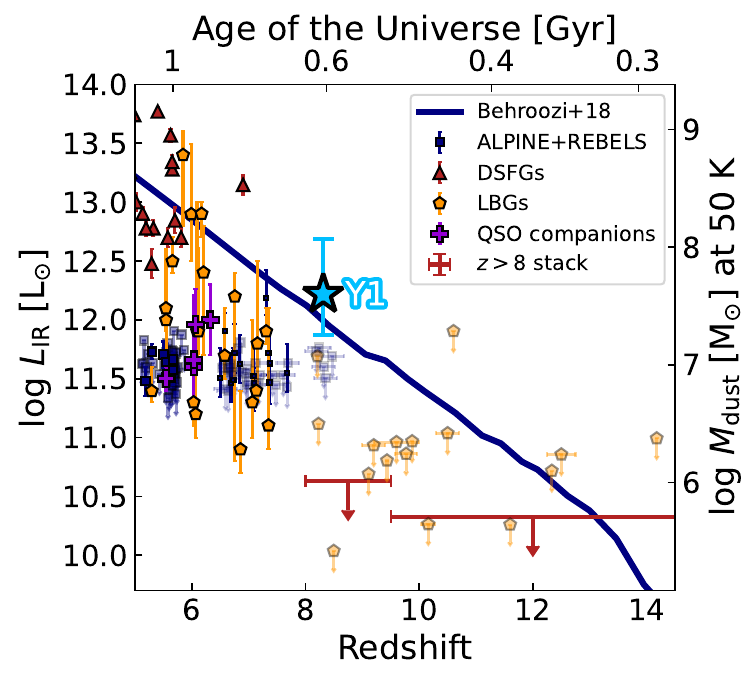}
    \caption{
    The infrared luminosity as a function of redshift of Y1 (\textit{light blue star}) is compared against those of individual and stacked $z > 8$ sources that remain undetected in deep sub-mm observations (\textit{dark red triangles}; Bakx et al. in prep.), as well as similar Lyman-break galaxies at $z = 5 - 15$ (\textit{orange pentagons}; \citealt{Laporte2019,faisst2020,Harikane2020,Sugahara2022ApJ...935..119S,Mitsuhashi2024Serenade} and Bakx et al. in prep.), UV-detected galaxies targeted in the ALMA Large Programs ALPINE and REBELS (\textit{blue}, with upper limits in \textit{light blue squares}; \citealt{lefevre2019,bethermin2020,Bouwens2022REBELS,Inami2022}), quasar-companion galaxies (\textit{purple pluses}; \citealt{venemans2020,Bakx2024QSOcompanions}), and dusty star-forming galaxies (\textit{red triangles}; \citealt{reuter20,ismail2023z,bendo2023}). 
    A model predicting the brightest dusty galaxy from a $0.2$ square degree survey size, assuming $T_{\rm d} = 50$K and a dust-to-stellar mass ratio of 0.01 (\textit{blue line}; \citealt{Behroozi2018}). Note that the warm dust temperature of Y1 causes its high luminosity with a more modest dust mass of $\sim 1.4 \times 10^6$M$_{\odot}$.
    }
    \label{fig:lvsz0}
\end{figure}

%
Y1 is a rest-frame UV-selected galaxy identified in a cluster-lensed field. While its initial identification was through bright UV-emission ($M_{\rm UV} = -20.9$), its modest dust mass obscures a large portion of the star formation, which in turn heats it up to warm temperatures. Figure~\ref{fig:lvsz0} shows the associated infrared luminosity of Y1, indicating its position as the {\color{referee} one of the infrared-brightest objects in the $z > 7$ Universe, and only galaxy with a direct dust detection above $z > 8$ to date}. Below, we will discuss Y1 in the context of dust in the high-redshift galaxy population.

ALMA has observed roughly ten sources beyond $z > 8.3$, but has not yet identified dust emission (a compilation and stack is provided in Bakx et al. in prep.). As such, Y1 represents the vanguard of dust in the Universe, and its high temperature immediately provides it a non-negligible luminosity. 
Lyman-break galaxies at $z = 5 - 8$ have been detected in their dust continuum \citep{Laporte2019,faisst2020,Harikane2020,Sugahara2022ApJ...935..119S,Mitsuhashi2024Serenade}, through for example the stunning discovery of dust at $z = 7.13$ in A1689-zD1 \citep{Watson2015,Knudsen2017,Bakx2021,Akins2022}. Since 2017, ALMA large programs have aimed to characterize the dust emission of UV-detected galaxies, with success rate between 30 and 50 per cent \citep{lefevre2019,bethermin2020,Bouwens2022REBELS,Inami2022,Algera2023}. It is important to realize that these surveys primarily search for dust surrounding the \ciil{} line, and consequently, this characterization would identify Y1 as a dust-poor system.\footnote{Although the galaxy nearest to Y1 in redshift with infrared emission, A2744-YD4, has a similar indication of warm dust \citep{Behrens2018,Laporte2017,Laporte2019,Algera2024}, its nature turned out to be more complicated \citep{Morishita2023,Hashimoto2023}. Thus, dust temperatures need to be carefully re-evaluated with higher-resolution ALMA observations} The UV-identified objects studied by the REBELS large program \citep{Bouwens2022REBELS} appear to have more massive dust reservoirs, and perhaps as a consequence of this, much lower dust temperatures \citep{Sommovigo2022REBELS,Ferrara2022REBELS,Algera2024}. 

At redshifts below 7, galaxies can be identified through their dust emission directly. Infrared surveys, starting with the mapping of the \textit{Hubble} Deep Field in 1997 \citep{Smail1997,Hughes1998}, in the sub-mm (e.g.,  \textit{Herschel Space Observatory} and JCMT/SCUBA-2) and mm (e.g., \textit{Planck}, {South Pole Telescope}, {Atacama Cosmology Telescope}) have identified over a million objects across fields with tens of square degrees to the entire sky. Consequently, these surveys readily detect dusty star-forming galaxies (DSFGs) with (apparent) luminosities beyond $> 10^{12.5}$\,L$_{\odot}$ \citep{berta2010,berta2011,Gruppioni2013}. The dust temperatures of these objects are low ($\sim 30 - 40$~K, albeit this $T_{\rm d}$ is potentially subject to optical depth effects; e.g., \citep{reuter20,Bendo2025}), but the vast dust reservoirs ($\sim 10^{9 - 10}$~M$_{\odot}$) of these systems \citep{ismail2023z,bendo2023} dominate the extragalactic infrared emission \citep{Zavala2021}.

The availability of \cii{} in favourable weather bands 6 and 7 -- and its important diagnostic value -- have incentivized observations of $z = 5 - 7.5$ quasars from large-area surveys. The combined survey area within the fields-of-view of these observations have resulted in tens of quasar-companion galaxies (QCGs) detected through line and/or continuum emission, which provided an important third probe between the UV- and sub-mm selection of galaxies in the early Universe \citep{Decarli2017,venemans2020,Bakx2024QSOcompanions,vanLeeuwen2024MNRAS.534.2062V}. Similar to previous interferometric blank field observations (the ALMA Lensing Cluster Survey and the ALMA Spectroscopic Survey in the \textit{Hubble} Ultra Deep Field), these aim to detect a large number of distant objects through their \cii{} emission. Interestingly, these galaxies appear to have low dust temperatures ($\lesssim 30$~K), large dust masses \citep[$\sim 10^8\,$M$_{\odot}$][]{Bakx2024QSOcompanions}, and high UV dust obscuration ($\gtrsim 93$~per cent; \citealt{vanLeeuwen2024MNRAS.534.2062V}), making them hard to include as a galaxy population from UV and even finding them directly in continuum-selected sub-mm observations. 

Y1 has a relatively low dust-to-stellar mass ratio of $1.4^{+1.3}_{-0.5} \times 10^{-3}$, in line with expected yields from supernovas and Asymptotic Giant Branch (AGB) stars \citep{Schneider2024Review}. The most massive of these stars, with $\sim 8$~M$_{\odot}$, require at least 30 to 100 million years to reach their AGB phase (\citealt{DwekCherchneff2011,Lugaro2012}). Assuming the majority of the dust in Y1 is produced through supernovae instead, based on the methodology in \cite{Michalowski2015}, we estimate a required supernova yield of $y = 0.07^{+0.06}_{-0.03}$~M$_{\odot}$/SN \citep[c.f.,][for an extended discussion]{Algera2024R25}. This is in line with models for supernova dust production ($\sim 0.2$ to 1~M$_{\odot}$; \citealt{Matsuura2011,Indebetouw2014,DeLooze2017,Niculescu-Duvaz2022}), as well as the expected destruction through reverse shocks that can destroy potentially up to 90~per cent of the dust formed \citep{Bianchi2007,Kirchschlager2019,Kirchschlager2024}. {\color{referee} Since the effect of dust destruction in Y1 (and elsewhere) are uncertain, it is reassuring to see the predicted dust yields below the theoretical estimates.}

Although the dust-to-stellar mass ratio is relatively low, an analytical one-dimensional dust evolution study predicts that this ratio still requires some time to achieve sufficient dust production from dust production sources \citep{Toyouchi2025}. Even in environments with enhanced destruction, fluid-dynamical models of dust production suggest that these dust-to-stellar mass ratios are not uncommon \citep{Esmerian2024}.
{\color{referee} \textit{Hubble} imaging revealed a young ($\sim 4$~Myr) stellar population. Due to the young age of this stellar population, models predict it is unable to produce the observed dust and oxygen emission in ALMA imaging. This implied the existence of a pre-existing massive, older stellar component \citep{Tamura2019}, based on the dust production model from \cite{Asano2013}. Updated spectral modeling of \textit{JWST} imaging find similarly-young stellar populations (\citealt{Ma2024ApJ...975...87M}; c.f., \citealt{Harshan2024}), which suggests that a pre-existing stellar population is responsible for the dust in Y1. }
If Y1 would be a progenitor of the most massive REBELS galaxies and QCGs, which have roughly an order of magnitude more massive stellar populations \citep{Algera2023}, there is a need for a substantial increase in the efficiency of dust production in its subsequent 200 million years, potentially due to the onset of rapid ISM dust growth \citep{Algera2025}.

A comparison to the galaxy evolution model of \citet{Behroozi2018} provides a measure of the maximum infrared luminosity for a galaxy detected in a field observed by deep optical observations, expected to hold true for most LBGs, ALPINE, REBELS, and the $z > 8$ LBGs, but not for DSFGs and QCGs identified from much larger area surveys. This comparison assumes a total survey size of $0.2$ square degrees roughly similar to the deep fields from \textit{Hubble} surveys \citep{oesch2016}. The highest stellar masses achieved in the model of \citet{Behroozi2018} are subsequently converted to an expected infrared luminosity (assuming $\beta_{\rm d} = 2$ and $T_{\rm d} = 50$~K) through a high stellar-to-dust mass ratio of 0.01. Y1 is in line with the highest expected infrared luminosity at this redshift, but primarily because of its high dust temperature, while its modest stellar mass instead lies well within the range expected for distant galaxies.

\subsection{Warm dust in the EoR}

Ever since the non-detection of dust continuum around the \cii{} emission of Y1, the suspicion of a warm dust reservoir in Y1 has existed. The observations reported in this paper now confirm the warm nature of the dust through an accurate census of the dust close to the peak of the dust continuum. 

Given the importance of warm dust in the early Universe, studies have reported both the direct fitting of the observed data and the implementation of dust models to limit the uncertainty on the dust temperature.
The first fit of the dust temperature \citep{Bakx2020CII} provided a $\chi^2$-based lower limit on the dust temperature of $> 80$~K. Using a more extensive fitting routine, \cite{Algera2024} reports on a lower limit at $3 
\sigma$ of $> 88$~K. We choose to take an optically-thin dust profile, as the estimated opacity at the observed wavelengths is negligible, assuming a dust mass limit of $\sim 1 \times{} 10^6$~M$_{\odot}$. 
Theoretical models of this source report similarly-warm dust temperatures, with the scaling relation modeling by \cite{Sommovigo2021} reporting a dust temperature of 82$^{+16}_{-19}$~K. Using the $91$~$\mu$m flux from \cite{Tamura2023}, a re-run of the modeling by \cite{Sommovigo2021} on Y1 finds similar results of $94^{+8}_{-11}$~K (see Appendix Figure~\ref{fig:Y1_newObs_Bakx}). Geometric arguments using radiative transfer under a clumpy sphere geometry in \cite{Inoue2020,Fudamoto2022DustTemperatures} predict a dust temperature of 95$^{+13}_{-16}$~K. These studies predict an associated demagnified infrared (8 to 1000~$\mu$m) luminosity of $\gtrsim 10^{12} L_{\odot}$, in line with a local ULIRG.

\begin{figure}
    \centering
    \includegraphics[width=\linewidth]{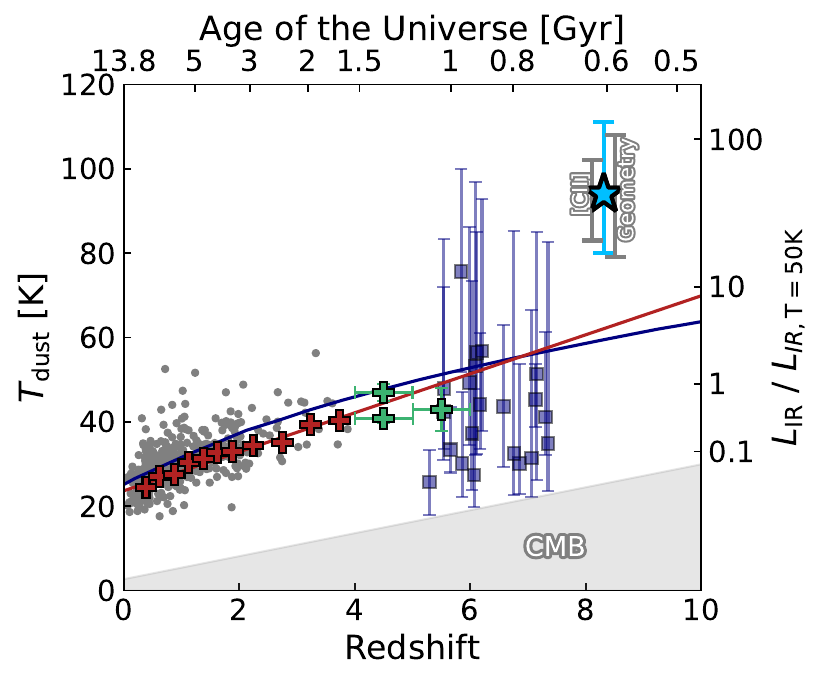}
    \caption{Dust temperature of galaxies and samples as a function of redshift. The dust temperature estimate for Y1 is shown in \textit{light blue}, and is compared to previous estimates based on \cii{} \citep{Sommovigo2021} and on a clumpy dust distribution geometric estimate \citep{Fudamoto2022DustTemperatures} in grey. Dust temperatures obtained for lower-redshift sources at $z < 4$ are shown in grey points \citep{Schreiber2018}, and for higher-redshift galaxies in blue squares (\citealt{faisst2020,Harikane2020,Sugahara2022ApJ...935..119S,Witstok2023Dust,Algera2024,Algera2024R25}). Dust temperature evolution based on stacked SEDs is shown in \textit{red and green pluses}, which are shown to increase linearly with redshift up to $z = 6$ \citep{Schreiber2018,bethermin2015,bethermin2020}. The best-fit from \citet{Schreiber2018} is shown in a \textit{red line}, while the average physically-motivated dust temperature relation from \citet{Sommovigo2022REBELS} is shown in a \textit{blue line}. Prior to the Band~9 data, only lower limits were available on Y1. The other axis indicates the change in infrared luminosity compared to 50~K, indicating the strong dependence of infrared luminosity on dust temperature, and the greyed-out region indicates the CMB temperature at a given redshift. }
    \label{fig:tvsz}
\end{figure}

Figure~\ref{fig:tvsz} shows the dust temperature of Y1 from a direct dust temperature fit from a modified black-body against those of the \cii{}-based dust temperature estimate from \cite{Sommovigo2021} and the clumpy dust geometry estimate from \cite{Inoue2020,Fudamoto2022DustTemperatures}. These are compared against $z < 4$ galaxy fits \citep{Schreiber2018} and a stack \citep{bethermin2015}. At $z > 4$, the dust temperatures come mostly from Lyman-break galaxies \citep{faisst2020,Harikane2020,Sugahara2022ApJ...935..119S,Witstok2023Dust,Algera2024,Algera2024R25}, as well as stacks of ALPINE data \citep{bethermin2020}. Two dust temperature scaling relations are shown, with one relation based on an empirical extrapolation of the $z < 4$ data \citep{Schreiber2018}. The other scaling relation is based on a physical model of dust attenuation in low metallicity and high optical-depth regions \citep{Sommovigo2022REBELS}. 

The dust temperature measures using the additional information, including \cii{} and the dust geometry, appear to provide a reasonable estimate, in line with the improved SED fit. Prior to the Band~9 data, only a lower limit at 80~K was available \citep{Bakx2020CII,Algera2024}. The short-wavelength observation now confirms the high dust temperature implied by the non-detection at 160~$\mu$m rest-frame, and places Y1 as one of the galaxies with the highest dust temperatures in the observed Universe, assuming an optically-thin modified black-body. {\color{referee}This high dust temperature is in line with the behavior of low-metallicity galaxies seen in the scaling relation in \cite{Sommovigo2022REBELS}. The metallicity measurements of Y1 find a $\sim 0.1-0.2 \, Z_{\odot}$ system \citep{Tamura2019,Harshan2024}, with modest variation in metallicity across the source \citep[][c.f.,]{Ma2024ApJ...975...87M}, in line with the observed high dust temperature of $\sim 90$~K. In these low-metallicity environments, dust is unable to self-shield, allowing for much higher dust temperatures than in more dust- and metal-rich galaxies, such as the higher-metallicity REBELS galaxies \citep{Algera2024,Algera2024R25,Rowland2025IFU}.}

\begin{figure}
    \centering
    \includegraphics[width=\linewidth]{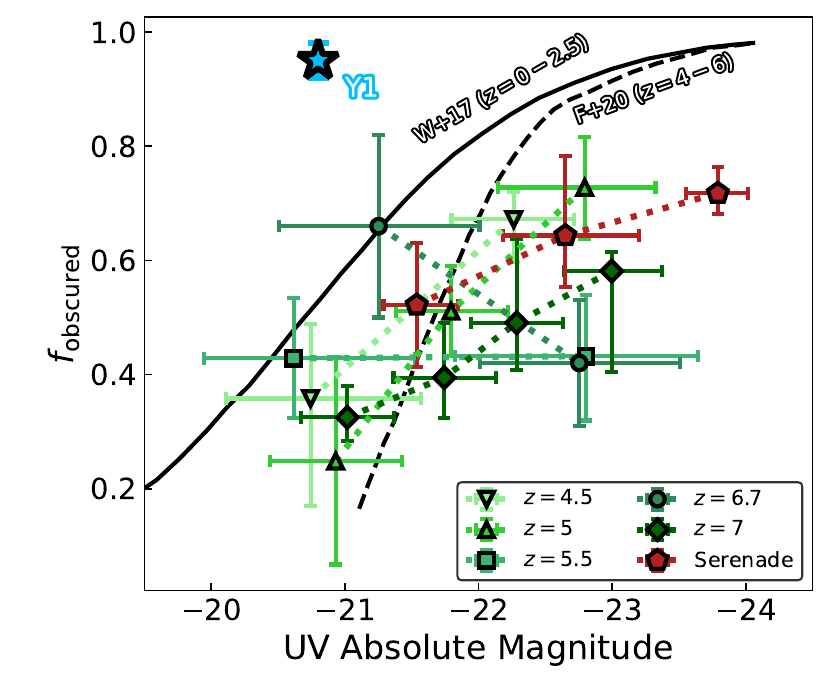}
    \caption{Obscured fraction of the star formation against the observed absolute UV magnitude ($M_{\rm UV}$-$f_{\rm obs}$). Y1 lies above the relation at a lower absolute UV-magnitude than previous stacked results.
The increasing hues of \textit{green downward triangles, triangles, squares, circles, diamonds}, and \textit{pentagons} are the results at $z\sim4.5$, 5, 5.5, 6.7 and 7 \citep{Fudamoto2020,Mitsuhashi2024CRISTAL,Bowler2024,Algera2023,Mitsuhashi2024Serenade}, respectively. The scaling relations from \citet{Whitaker2017,Fudamoto2020} are shown in solid and dashed lines, respectively.}
    \label{fig:muv_fobsc}
\end{figure}

As a consequence of the high dust temperature, the infrared luminosity is $\sim 35$ times higher than the typically-assumed 50~K luminosity on a single continuum datapoint\citep[e.g.,][]{Bouwens2020}. The unobscured star-formation rate of Y1 is $14 \pm 1$~M$_{\odot}$/yr for a system with a total UV luminosity of $(4.00 \pm 0.18) \times{} 10^{10}$~L$_{\odot}$ \citep{Hashimoto2019,Ma2024ApJ...975...87M}. Meanwhile, assuming an IR luminosity-to-SFR conversion factor of $1.73 \times{} 10^{-10}$ M$_{\odot}$yr$^{-1}/$L$_{\odot}$ (valid for a Salpeter $1-100$~M$_{\odot}$ IMF), the obscured star-formation rate of Y1 is 173~M$_{\odot}$/yr. Consequently, the obscured fraction of star-formation of Y1 is high, at $93 \pm 5$~per cent. 
Figure~\ref{fig:muv_fobsc} shows the high obscured fraction of Y1 against the average obscuration fractions of galaxies \citep[][]{Fudamoto2020,Mitsuhashi2024CRISTAL,Bowler2024,Algera2023,Mitsuhashi2024Serenade} and scaling relations \citep{Whitaker2017,Fudamoto2020} at lower redshifts. Y1 lies above these relations, indicating it has a high obscured fraction of star-formation when compared to samples of galaxies at lower redshift.

The high luminosity of Y1 originates from an infrared-dominated spectrum, and places it in the ULIRG (or ULIRG-like) regime. At cosmic noon, ULIRGs are typically found to be dust-obscured galaxies with cold dust \citep[e.g.,][]{bendo2023,Bendo2025} and high dust-to-gas ratios \citep[e.g.,][]{Hagimoto2023}. Alternative to these colder galaxies, 
dust temperatures of about 90\,K are observed in other high-$z$ sources and in some cases associated to AGN host galaxies. For example, WISE identified a set of hot-dust obscured galaxies (affably coined HotDOGs; \citealt{Tsai2015}), where the bolometric luminosity is dominated by a hot dust component likely powered by massive Active Galactic Nuclei (AGN; \citealt{FA2025HotDog}). Submm bright hot-DOG host galaxies have been found harbour SMG-like starbursts with temperatures of $60-100$\,K (e.g., \citealt{Fan2016}).
While the optical emission lines indicate a potential narrow-line AGN in Y1, the presence of two dust emission peaks in the high-resolution Band~7 data \citep{Tamura2023} implies that there is not a single central heating source of the dust. Both of these components are likely originating from separate warm dust components, as otherwise one of the two faint components would be detected in the Band~5 data (see Astles et al. in prep.). 
The presence of multiple warm dust components could in part be due to heating by this narrow-line AGN \citep[e.g.,][]{Tsukui2023,FA2025HotDog}, but is more likely to originate from stellar emission, in line with models of compact, young star-forming regions enshrouded in molecular clouds \citep{Behrens2018,Pallottini2022}. 

A discrepancy between the obscured and unobscured cosmic star-formation rate density has become apparent in the early Universe (e.g., \citealt{Casey2018}). Recent works using deep sub-mm (ALMA) and radio (Very Large Array) \citep[e.g.,][]{Novak2017,Gruppioni2020,Talia2021} indicate a high SFRD with little evolution out to redshifts beyond $z > 3$. 
The uncertainty in the obscured SFRD at high redshift is due to the lack of deep sub-mm surveys capable of characterizing the dust temperature at high-$z$. The purported hot dust temperature expected for low-metallicity systems \citep{Sommovigo2022REBELS} could produce a population of low-metallicity ($< 0.2 Z_{\odot}$) young galaxies with small but excessively hot dust bodies that dominate the cosmic SFRD. Missed in longer-wavelength surveys \citep{Zavala2021,Long2024MORA}, the high obscuration fraction of Y1-like galaxies ($93 \pm 6$~per cent) would be enough to flatten the $<1$~dex. drop in cosmic SFRD between $z = 3$ to 8. Meanwhile, the warm dusty phase is likely very short. 
The high specific star-formation rate of 187~Gyr$^{-1}$ indicates the likely short-lived nature of these systems, which is in line with young star-bursting systems.

Similarly, the contrast between the \cii{} and \oiii{} emission implies an ionized gas region outflowing from Y1 with $\sim 100$~km/s. This outflowing cloud (about one-third of all \oiii{} luminosity) traces the UV-emission (Bakx et al. in prep.). As the star-formation rate is $> 150$~M$_{\odot}$/yr within this $\sim 1$~kpc$^2$ system, and the small \textit{JWST}-estimated stellar mass ($\sim 10^{9}$~M$_{\odot}$) is unlikely to keep this gas contained (\citealt{Andrews2011}; c.f., \citealt{Kohandel2025}). 

Such episodic star-formation in the early Universe can drive fast metal-rich winds into the circum-galactic medium (CGM).
Indications of star-formation driven winds are fundamental ingredients of hydrodynamical (e.g., \citealt{Arata2019}), semi-analytical (e.g., \citealt{Yung2019}) and analytical \citep{Ferrara2023} models, as well as crucial for explaining dust \citep{Bakx2024QSOcompanions}, the scatter in the mass-metallicity relation \citep{Pallottini2025}, and emission line observations \citep{Carniani2024} in low-mass galaxies in the early Universe. The $ \sim 100$~km/s velocity offset seen through combined \cii{} and \oiii{} emission is in line with the analytically-predicted velocities of radiation-driven winds \citep{Ferrara2023,Ferrara2024}. 


%

\subsection{IRX-$\beta_{\rm UV}$ relation in the early Universe}
Dust obscures our UV and optical view of star formation by preferentially obscuring the blue components of the stellar emission. The subsequent heating of the dust results in bright (sub)mm emission from galaxies. The dust obscuration can be quantified both through the difference between obscured and unobscured emission, as well as through the reddening of the rest-frame UV emission. 
Common dust-attenuation curves can be used to estimate the sub-mm emission and associated dust-obscured star-formation rate directly from the UV luminosity ($L_{\rm UV}$) and slopes ($\beta_{\rm UV}$) that identify $z > 7$ galaxies through the so-called infrared excess ($\rm IRX = \rm \log_{10}$~$L_{\rm IR}/L_{\rm UV}$). This correlation between $\beta_{\rm UV}$ and IRX balances the absorbed UV emission that is subsequently re-emitted in the sub-mm regime. {\color{referee}However, a correlation between IRX and $\beta_{\rm UV}$ can only occur if the UV and dust-emitting regions are cospatial \citep{Howell2010,Faisst2017}, as otherwise the UV emission can be dominated by bright unobscured stars that outshine the contribution of the obscured stars which heat the dust, further exacerbated by complex geometries \citep{Vijayan2024}. Stark deviations from this IRX-$\beta_{\rm UV}$ relation are evidenced by the existence of optically-dark galaxies (e.g., \citealt{Wang2019OptDark,Casey2019,Zavala2021,Fudamoto2021OpticallyDark}) and optically-detected galaxies (e.g., \citealt{Carniani2017,Bowler2022,Bowler2024,Inami2022}) with significant spatial offset between their ALMA and \textit{HST} data,} as well as the identification of dusty star-forming galaxies before them \citep{Smail1997,Hughes1998}. Observations in the UV can thus not make up for direct infrared observations, as also supported by theoretical studies and simulations (e.g., \citealt{Ferrara2022REBELS}).

\begin{figure}
    \centering
    \includegraphics[width=\linewidth]{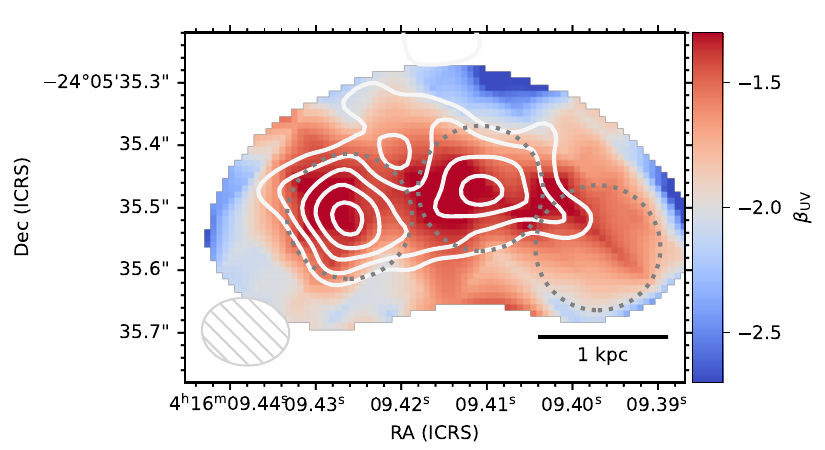}
    \caption{The spatially-resolved UV-continuum slope ($\beta_{\rm UV}$) relative to the Band~7 dust continuum (\textit{white contours}). The peak positions of the ALMA-based dust continuum is cospatial to relatively high, red values of $\beta_{\rm UV}$. This shows that the dust obscuration of stellar emission in both sub-mm and optical colours originate from similar regions. The small circles indicate the regions where the IRX-$\beta_{\rm UV}$ is evaluated in a resolved fashion in Figure~\ref{fig:irxbeta}. }
    \label{fig:betauv}
\end{figure}

The combined observations at $\sim 0.1$~arcsec resolution from \textit{JWST} and ALMA Band~7 allow us to compare this reddening to the total emitted (sub)mm emission. Figure~\ref{fig:betauv} shows the resolved $\beta_{\rm UV}$ map from \textit{JWST} at $\sim 200$~pc resolution. 
We use the filters F150W, F200W and F277W to produce the resolved $\beta_{\rm UV}$ map, which are reprojected and smoothed to the same common point-spread function (PSF), using the empirically-derived PSFs for the \textit{JWST} filters. The observed wavelengths are 15010~\AA{} for the F150W, 19880~\AA{} for the F200W and 27760~\AA{} for the F277W filter. These images are astrometrically verified against Gaia stars and previous \textit{Hubble} imaging \citep{Tamura2019}, and are shown in Appendix Figure~\ref{fig:JWSTcompilation}. A combined RGB-image of the smoothed \textit{JWST} imaging can be found in Appendix Figure~\ref{fig:rgbJWST}. Per pixel, a single $\beta_{\rm UV}$ is fit to the fluxes and error maps across three wavelengths, overcoming the uncertainties found to occur in previous measurements from colour-colour slopes \citep{Cullen2024}. 

In an effort to exclude line confusion, we explore the available NIRSpec Integral Field Unit (JWST PID 1208; P.I. Chris Willott) and NIRSpec Multi-Object Spectroscopy \citep{Harshan2024}. While the stellar continuum is visible in the IFU and PRISM spectroscopy, no bright line emission is visible nor expected in within the expected photometric filters F150W, F200W and F277W. Consequently, we predict the effect of line contamination to the $\beta_{\rm UV}$ image to be negligible. Although beyond the scope of this paper, we note that future work could continue this investigation into the resolved attenuation through direct measurement of the Balmer line attenuation, which is less affected by stellar age and the assumption on the intrinsic $\beta_{\rm UV,0}$ slope to calculate the $A_V$.

The two dust continuum peaks are close to the peaks in the reddest part of the $\beta_{\rm UV}$ map, indicating that the dust emission appears cospatial with $\beta_{\rm UV} = -1.2$ to $-1.7$ \citep{Tamura2019,Harshan2024}. Meanwhile, 
the spur-like feature to the south-west part of Y1 is co-spatial to a red component with $\beta_{\rm UV} \approx -1.6$, which is brightly seen through UV imaging but not in dust continuum imaging of Y1.

\begin{figure}
    \centering
    \includegraphics[width=\linewidth]{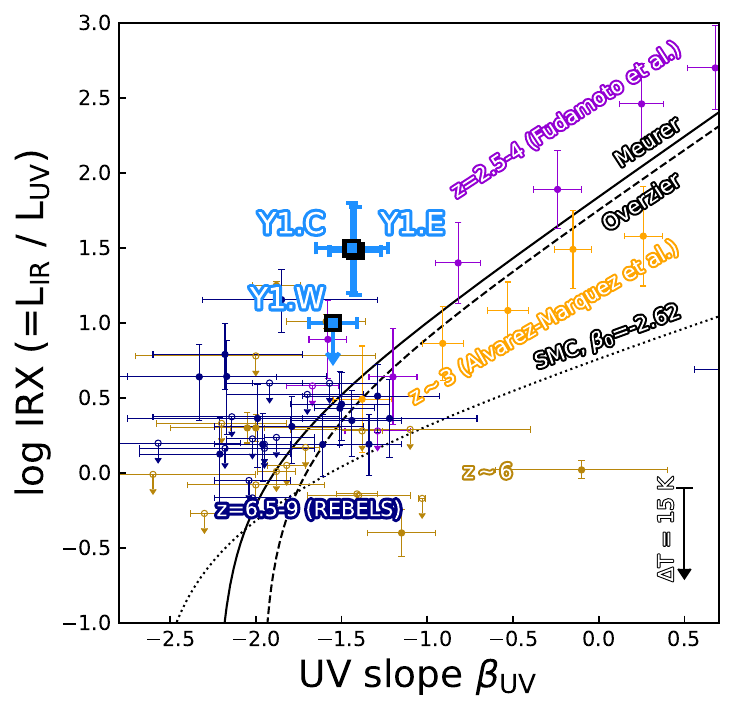}
    \caption{The infrared excess of Y1 is evaluated at three locations and compared against the observed UV continuum slope ($\beta_{\rm UV}$). These are compared to 
previous measurements of the galaxies at $z=2.5$--4 \citep[][; purple]{Fudamoto2020cosmos}, $z\sim3$ \citep[][orange]{AlvarezMarquez2019}, $z\sim4.5$ and $z\sim5.5$ \citep[][blue circles]{Fudamoto2020}, and $z\sim6$ \citep[][golden circles]{Bowler2024}.
These are further compared against the best-fit relations from local \citep{Prevot1984A&A...132..389P,Meurer1999ApJ...521...64M,Calzetti2000,Gordon2003,Overzier2011ApJ...726L...7O,Takeuchi2012} and high-$z$ galaxies \citep[e.g.,][]{AlvarezMarquez2016,Reddy2018,Fudamoto2020}. {\color{referee} A spatial variation in dust temperature of 15~K is indicated by an arrow in the bottom-right of the panel.} }
    \label{fig:irxbeta}
\end{figure}

Figure~\ref{fig:irxbeta} quantifies the UV-continuum slope and the infrared-to-UV emission through the so-called IRX-$\beta$ relation. The $\beta_{\rm UV}$ is extracted at the peak dust emission, as well as at the Western component that does not show indications of dust emission in 0.2 arcsecond apertures. The resolved infrared and UV luminosity are derived by their respective contribution to the total flux in ALMA Band~7 and F200W NIRCam imaging. 

The two dust-emitting regions lie above the observed scaling relations, in line with the \textit{dust screen with holes} scenario \citep{Popping2017}. CLOUDY modeling of this galaxy \citep{Hagimoto2025}, including its spectral lines and dust emission, predicts a low covering fraction of neutral gas in this galaxy of 25~per cent, suggesting 75~per cent of the ionized gas region is exposed to intercloud space. This is similar to dwarf galaxies seen in the local Universe \citep{Cormier2019}. Indeed, while the dust and $\beta_{\rm UV}$ values imply that dust absorption coincide, the deviation from the IRX-$\beta_{\rm UV}$ indicates that the dust attenuation of stars occurs on scales smaller than those probed by \textit{JWST} and ALMA at $\sim 200$~pc \citep{Kruijssen2014,Kruissen2019Natur.569..519K,DiMascia2025}. {\color{referee}Potentially, temperature variation across Y1 can affect the resolved IRX measure. The dust temperature in both components is likely still high, as otherwise it would be detected in the resolved 160~$\mu$m dust emission map (see Bakx et al. in prep.). Figure~\ref{fig:irxbeta} indicates the effect of a large dust-temperature variation of $\sim 15$~K \citep{Akins2022,Bowler2022,Tsukui2023,FA2025HotDog}. In this extreme case, this could reduce the IRX-$\beta_{\rm UV}$ discrepancy of one of the components of Y1, while exacerbating the discrepancy for the other. A more detailed analysis of the spatial variation of dust temperature is beyond the scope of this work.}

This measure can be further quantified through the so-called molecular index, defined as $I_m = (F_{158}/F_{1500})/(\beta_{\rm UV} + 2.6)$ \citep{Ferrara2022REBELS}, that compares the fluxes $F$ at the sub-mm (158~$\mu$m) against those in the rest-frame optical (1500~\AA), and compares this against the steepness of the UV continuum slope. For Y1, the observed $F_{1500}$ flux across the source is $0.124 \pm 0.03$~$\mu$Jy \citep{Ma2024ApJ...975...87M} based on the nearest \textit{JWST} filter F150W, with the $F_{158} = 12 \pm 6$~$\mu$Jy. The resulting $I_m$ ranges from 70 to 110, taking the spread in values for $\beta_{\rm UV} = -1.2$ to $-1.7$ seen across the source. Not exceeding the threshold for strongly segregated obscured and unobscured regions of $I_m > 1120$, the ISM is likely a mixture of mixed and separated \citep{Ferrara2022REBELS,Sommovigo2022}. 
This is in line with the previous finding that the bulk emission of stars and dust emission is unassociated on the scales of  $\sim 500$~pc \citep{Tamura2023}, as well as its low neutral gas covering fraction of $\sim 25$~per cent \citep{Hagimoto2025}. In this scenario, strong stellar feedback from young stellar populations exposes them to low-attenuation sightlines, where they outshine the redder, attenuated emission from obscured stellar regions \citep{Arata2019}.

\section{Conclusions}
\label{sec:conclusions}

The Band 9 observations reported in this paper confirm that Y1 is a ULIRG at redshift 8.3 by tracing its mm and sub-mm dust spectrum. Although this source was initially detected with a similar sub-mm emission as $z = 7 - 8$ galaxies \citep{Tamura2019}, its uniquely high dust temperature (91$^{+62}_{-35}$~K) results in a ULIRG-like infrared luminosity, with a high obscured fraction of star-formation ($\sim 93$~per cent). The short-wavelength observations thus confirm the initial suspicions of a high dust temperature, raised by the non-detection of the Band~5 emission \citep{Bakx2020CII}. The to-date furthest direct detection of dust emission thus originates from a stand-out system with a high temperature, luminosity and obscuration, with only a modest dust mass in line with typical models of supernova and AGB dust production. When spatially comparing the resolved Band~7 dust emission to the dust-attenuated UV continuum slope ($\beta_{\rm UV}$), the dust-emitting regions lie cospatial to optically-obscured regions. However, the elevated position of Y1 in the IRX-$\beta_{\rm UV}$ plot indicates the existence of obscured stellar components that are missed by the UV observations of this ULIRG. This suggests that dust is only partially covering the UV-emitting regions, and/or that dust and UV-emitting regions are dissociated on scales smaller than 200~pc, i.e., those probed by \textit{HST}, \textit{JWST} and ALMA.


\section*{Acknowledgements}
{\color{referee}The authors kindly thank the anonymous referee for their insightful comments and suggestions to improve this manuscript.}
This paper makes use of the following ALMA data: ADS/JAO.ALMA 
2013.1.00999.S,
2017.1.00225.S,
2019.1.00343.S,
2019.1.01350.S,
2021.1.00075.S,
2022.1.01356.S,
2024.1.00537.S, and
2024.1.00995.S.
Financial support from the Knut and Alice Wallenberg foundation is gratefully acknowledged through grant no. KAW 2020.0081. KK acknowledges support from the Knut and Alice Wallenberg Foundation and the ERC Synergy Grant "RECAP" (grant 101166930).
This work was supported by NAOJ ALMA Scientific Research Grant Numbers 2018-09B and JSPS KAKENHI No. 17H06130, 22H04939, 23H000131, and 25H00671. J.M. and E.I. gratefully acknowledge financial support from ANID - MILENIO - NCN2024\_112. E.I. also gratefully acknowledges financial support from ANID FONDECYT Regular 1221846.
AF acknowledges support from the ERC Advanced Grant INTERSTELLAR H2020/740120.

\section*{Data Availability}
The datasets generated during and/or analysed during the current study are available from the corresponding author on reasonable request. 
The codes used to reduce and analyse the ALMA data are publicly available.



\bibliographystyle{mnras}
\bibliography{1_example} 

\appendix

\section{Band 9 observing conditions}
\label{sec:whyband9}
Table~\ref{tab:observationdetails} shows the parameters of the ALMA Band 9 observations. 
Band~9 was taken to characterize the dust temperature through a series of simulated observations of Y1 with potential dust temperatures between 60 and 120~K. Given the existing dust detections and upper-limits in Bands~5, 7 and 8, much deeper Band 8 (10 hours at 410, 461 or 487 GHz) is unable to provide a robust dust temperature estimate, as it does not probe close enough to the peak to constrain the dust temperature accurately. Meanwhile four hours of Band 10 (at 870 GHz) is not sensitive enough to detect a dusty body at $\approx 70$~K with high significance. This modeling further shows that Band~9 observations at 671~GHz are preferred relative to the slightly-higher frequency option at 690~GHz, where the latter option has a slightly higher optical opacity.

\begin{table}
    \centering
    \caption{Parameters of the ALMA Band 9 observations}
    \label{tab:observationdetails}
    \begin{tabular}{lccccccc} \hline
UTC end time          & Baseline length & N$_{\rm ant}$  & T$_{\rm int}$ & PWV \\
  & [m] 			 & 			 	     & [min] &  [mm] \\ \hline
2024-12-20 04:37:30 & 14 -- 499 & 45 & 74.11 & 0.42 \\
2024-12-13 06:04:48 & 14 -- 313 & 44 & 73.10 & 0.59 \\
2024-12-10 05:13:50 & 15 -- 313 & 45 & 73.27 & 0.59 \\
2024-11-09 05:06:37 & 14 -- 313 & 43 & 74.42 & 0.34 \\
 \hline
    \end{tabular}
    \raggedright \justify \vspace{-0.2cm}
\textbf{Notes:} 
Col. 1: UTC end time of the observations as $[$YYYY-MM-DD hh:mm:ss$]$.
Col. 2: Length between the nearest and furthest antennae.
Col. 3: Number of antennae participating in the observations.
Col. 4: The total observation time including overheads.
Col. 5: The precipitable water vapor during the observations.
\end{table}

\section{[CII]-based dust temperature estimate}

\begin{figure}
    \centering
    \includegraphics[width=\linewidth]{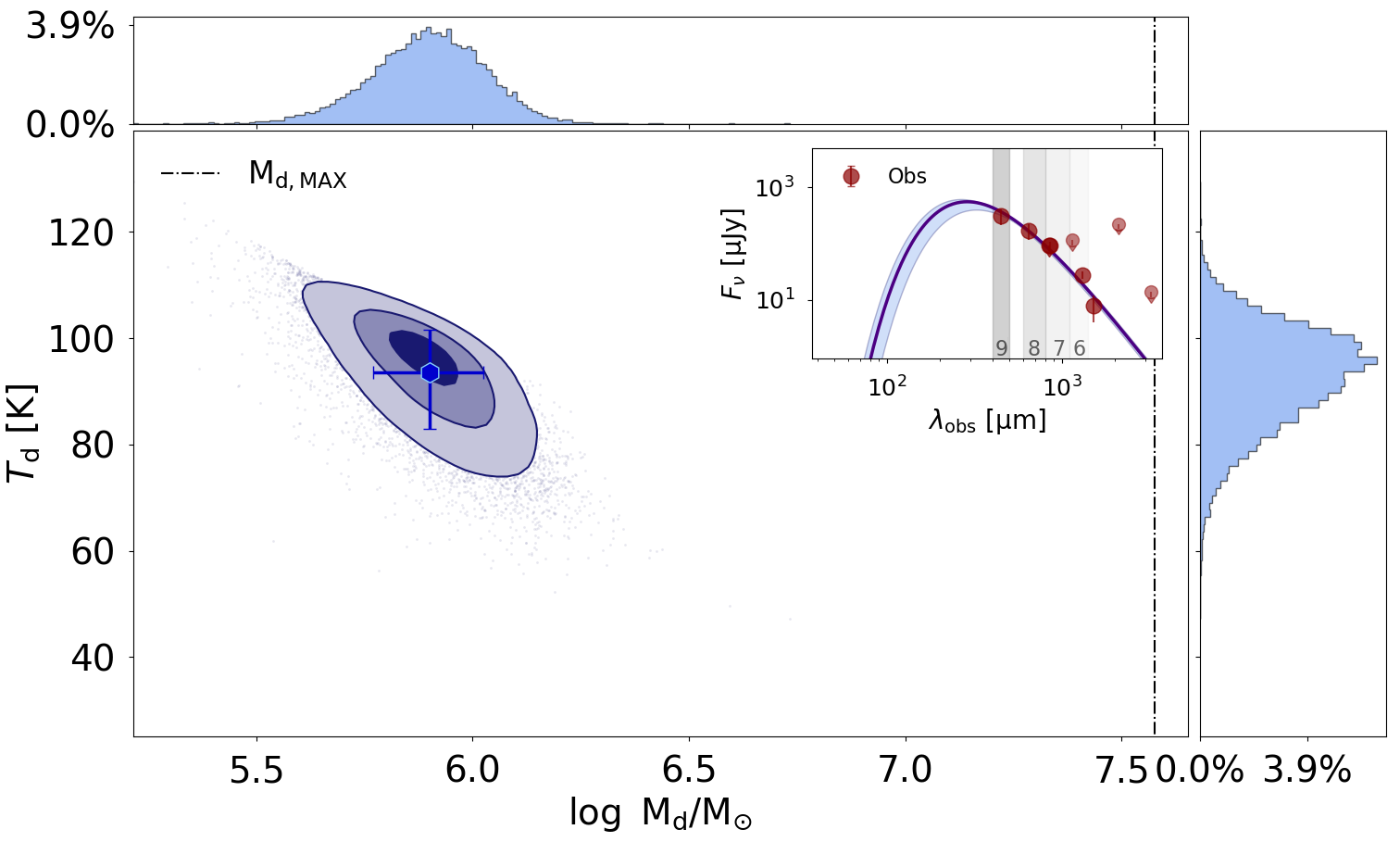}
    \caption{Recovered distribution of $T_{d, [\rm C_{II}]}$ as a function of dust mass for Y1 using the updated fluxes detailed in Table~\ref{tab:data}. The resulting dust temperature ($94^{+8}_{-11}$~K) and dust mass ($8.0_{-2.0}^{+2.6} \times{} 10^5$~M$_{\odot}$) broadly agree with the direct fitting results. The contours show the 16$^{\rm th}$, 50$^{\rm th}$ and 84$^{\rm th}$ percentiles of the distribution, with the median value represented by the purple square. The best-fit spectrum, also shown in \textit{black dashed lines} in  Figure~\ref{fig:SEDfit}, is shown in the inset, together with all data and upper limits.}
    \label{fig:Y1_newObs_Bakx}
\end{figure}

Using the methodology detailed in \cite{Sommovigo2021}, a refit of Y1 is made using the Band~7 continuum flux of $137 \pm 26$~$\mu$Jy at 353~GHz. Using the updated stellar mass from JWST imaging of 10$^9$~M$_{\odot}$, as well as a characteristic \cii{} size between once and twice that of the stellar emission (Bakx et al. in prep.), a dust fit is performed in Figure~\ref{fig:Y1_newObs_Bakx}. The resulting dust temperature is $94^{+8}_{-11}$~K, with a total dust mass of $8.0_{-2.0}^{+2.6} \times{} 10^5$~M$_{\odot}$, similar to those from the direct SED fit shown in Figure~\ref{fig:SEDfit}.


\section{JWST data for $\beta_{\rm UV}$ calculation}
\begin{figure*}
    \centering
    \includegraphics[width=\linewidth]{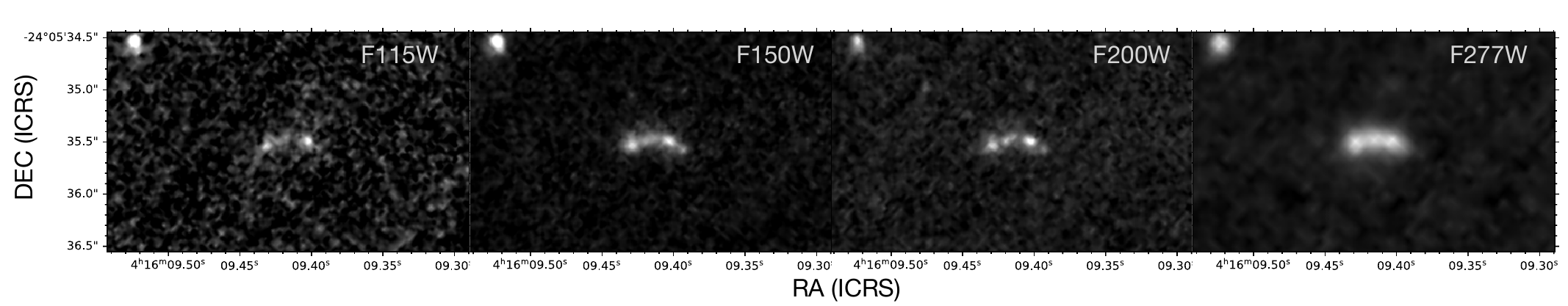}
    \caption{The \textit{JWST} photometric images in filters F115W, F150W, F200W and F277W. The latter three are used for the $\beta_{\rm UV}$ analysis, after smoothing everything to the resolution of F277W. }
    \label{fig:JWSTcompilation}
\end{figure*}

\begin{figure}
    \centering
    \includegraphics[width=\linewidth]{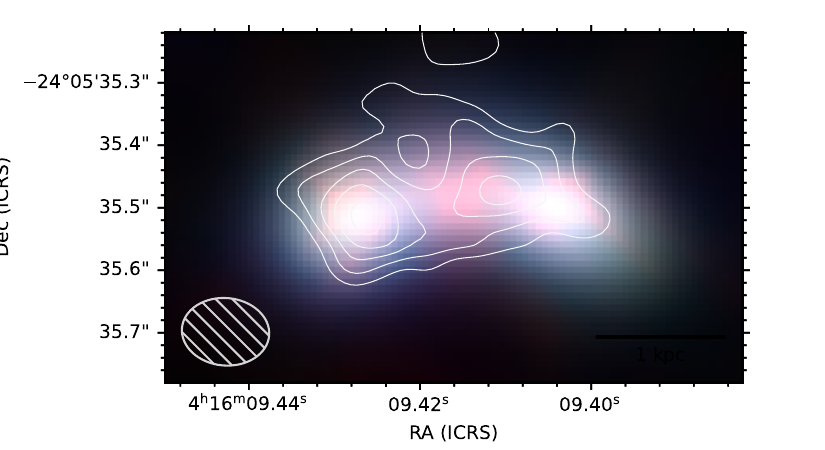}
    \caption{The combined \textit{JWST} photometric images in filters F150W (b), F200W (g) and F277W (r). }
    \label{fig:rgbJWST}
\end{figure}

Figure~\ref{fig:JWSTcompilation} shows the compilation of \textit{JWST} data used for calculating the spatially-resolved $\beta_{\rm UV}$ values. The combined red-green-blue (rgb) image of Y1 is shown in Figure~\ref{fig:rgbJWST}.

\bsp	
\label{lastpage}
\end{document}